\documentstyle[12pt]{article}
\newcommand{\NN}{
\begin{picture}(11,3)(-2,-2)
\put(1,-12){$\tilde{}$}
\put(-2,-2){$N$}
\end{picture}}

\begin{document}
\baselineskip=22pt plus 0.2pt minus 0.2pt
\lineskip=22pt plus 0.2pt minus 0.2pt
\begin{center}
\vspace*{1cm}
\LARGE
A Comment on the Degrees of Freedom \\
in the Ashtekar Formulation for 2+1 Gravity\\

\vspace*{1.5cm}

\large                                                                        

J.\ Fernando\ Barbero\ G$^*$,\\and\\
Madhavan Varadarajan$^{**}$

\vspace*{1.5cm}

\normalsize
$^*$Escuela Superior de Ingenier\'{\i}a Industrial,\\
Universidad Europea de Madrid\\
Urb. El Bosque, C/ Tajo s/n\\
Villaviciosa de Od\'on, Madrid, 28670\\
Spain\\\hspace{5mm}\\
$^{**}$ Raman Research Institute\\
Sir C. V. Raman Avenue\\
Bangalore 560 080, India\\

\vspace{.3in}
July 14, 1999\\
\vspace{.3in}
ABSTRACT
\end{center}

We show that the recent claim that the 2+1 dimensional Ashtekar 
formulation for General Relativity has a finite number of 
physical
degrees of freedom is not correct.

\vspace*{1cm}
\noindent PACS number(s): 04.20.Cv, 04.20.Fy

\pagebreak

\setcounter{page}{1}        

In a recent paper \cite{MM1} Manojlovi\'c and Mikovi\'c claim 
that the number of degrees of freedom in Ashtekar's formulation 
for 2+1 dimensional General Relativity is finite at variance 
with previous results of the authors \cite{VB}. 
We stand by the results of 
that paper wherein we were
able to prove that, in spite of having the same number of first
class constraints as phase space variables per point, the 
number  of degrees of freedom in this formulation is {\it 
infinite}. In this comment we show that \cite{MM1} is incorrect 
on several counts.\\

\noindent {\bf (i)} The  statement appearing in 
page 3034:

\vspace{.5cm}

\parbox{13.5cm}{
{\it \noindent ``..., by performing a gauge transformation on 
a null connection, one can always reach a non-null connection
..."}
}

\vspace{.5cm}
\noindent is not true. This statement is `proved' by (25) of 
\cite{MM1}.
However, (25) of that paper is incorrect because the gauge 
transformations 
of the connection that the authors have used are wrong. 
Specifically, 
eq.(17) should say
$$\delta A_1=-\frac{d\epsilon^1}{d\theta}+
(E_2 f_3-E_3 f_2)\epsilon^3$$
and eq.(19) should be
$$\delta A_3=-A_2\epsilon^1+f_3\epsilon^2+E_1f_2\epsilon^3$$
The origin of the error in the last equation can be traced back 
to the 
use of an incorrect symplectic structure to derive it from the 
constraints. 
$(E^a_I {\dot A}_a^I)$ evaluated on (9), (10) of \cite{MM1}
is $E_1{\dot A}_1+E_2{\dot A}_2-E_3{\dot A}_3$ {\em not} 
$E_1{\dot A}_1+E_2{\dot A}_2+ E_3{\dot A}_3$ as in (14) of 
\cite{MM1}.
This is because $I$ is an SO(2,1) index and $t_I t^I=-1$ 
{\em not} +1.
Correction of these errors leads to the following version 
of equation (25)
\begin{eqnarray}
\delta (f_2 \pm f_3) & = & 
 -\epsilon^1(f_3 \pm f_2) + [\epsilon^2 (f_2 \pm f_3)]^{\prime}
 - \epsilon^2 A_1 (f_3 \pm f_2) \label{eq:correct}\\
&  & + [\epsilon^3 E_1(f_3 \pm f_2)]^\prime - 
\epsilon^3 A_1 E_1(f_2 \pm f_3)
-\epsilon^3 (E_2 f_3-E_3 f_2)(A_3\pm A_2) ,\nonumber
\end{eqnarray}
where
$$f_2\equiv\frac{dA_2}{d\theta}-A_1 A_3,
\hspace{5mm}f_3\equiv\frac{dA_3}{d\theta}-A_1 A_2.$$
>From (\ref{eq:correct}) it is straightforward to obtain
\begin{eqnarray}
\delta(f_2^2-f_3^2)& = & 2 (\epsilon^{2})^{\prime}(f_2^2-f_3^2)+
\epsilon^2(f_2^2-f_3^2)^{\prime}+\epsilon^3 
\left[2A_1 E_1 (f_3^2-f_2^2)+\right.
\nonumber\\
& & \left. 2E_1(f_2 f_3^{\prime}-f_3 f_2^{\prime})+
2(E_2f_3-E_3 f_2)(A_2 f_3-A_3 f_2)\right]\label{neweq}
\end{eqnarray}
At an interior point in a null patch, we have $f_2^2-f_3^2=0$,
$(f_2^2-f_3^2)^{\prime}=0$, and $f_2=\pm f_3$ so, modulo the 
constraints,
\footnote{Note that for flat curvatures, $f_2=f_3=0$ and
 the last term in (\ref{neweq}) vanishes. For non flat,
 null curvatures
$f_2=\pm f_3 \neq 0$ and the `vector constraint', (12), 
of \cite{MM1}
implies $E_2=\pm E_3$. Thus $E_2f_3-E_3 f_2$ vanishes 
and so does the last
term in (\ref{neweq}).}
(\ref{neweq}) gives $\delta(f_2^2-f_3^2)=0$. 
Clearly, this shows that the statement on pg 3034 of \cite{MM1}, 
referred to 
above  is incorrect.\\

\noindent{\bf (ii)} The authors claim that the holonomies around
loops in  flat patches
are not gauge invariant objects. 
This claim is not correct. 
Note that we did 
{\em not}
specify the location of the loops in fixed coordinates but
 demanded that 
they be associated with flat patches 
(since the connection is flat the holonomy is independent of 
the 
location of the loop within a flat patch). Hence, the loops 
are moved by the 
same diffeomorphism which moves the flat patches under 
evolution. Thus these
observables are 2d diffeomorphism
 invariant besides obviously being SO(2,1) gauge 
invariant. Since these are all the gauge generated by the 
constraints on the
connection part of the patch data, 
the holonomies are gauge invariant observables. Indeed,
it is easily seen that (denoting the holonomy around the 
loop $\gamma$ in a
flat patch by $H_\gamma$),
\begin{eqnarray}
\{H_\gamma, {\cal D}_a E^a_I\}&=&0  \\
\{H_\gamma, E^a_I F_{ab}^I\}&= &\{H_\gamma,E^{a}_I\}F_{ab}^I=0 
\label{eq:pb1}\\
\{H_\gamma, \epsilon^{IJK}E^a_IE^b_JF_{abK}\} & = & 
\{H_\gamma, \epsilon^{IJK}E^a_IE^b_J\} F_{abK} =0. 
\label{eq:pb2}
\end{eqnarray}
where we used $F_{ab}^I=0$ in (\ref{eq:pb1}),(\ref{eq:pb2}).
These Poisson bracket relations are {\em independent} of the 
degeneracy 
of the metric. 

\vspace{3mm}

\noindent{\bf (iii)} The authors have misinterpreted our display 
of initial
data in \cite{VB} as a gauge fixing condition. We have not fixed 
any gauge 
in that work! We summarize the arguments of \cite{VB} to 
emphasize this point.
We studied  the 
action of the gauge transformations generated by the constraints
on certain types of initial
data. 
Specifically, we analysed
data characterized by N (an arbitrary positive integer) non 
simply connected 
annuli on the torus where the curvature was null and with zero 
curvature 
elsewhere (the torus is a product of 2 circles and can be 
coordinatized by
2 angles $(\theta, \phi)$; each annulus lies in some interval
 of $\theta$
with $\phi$ going through its full range). 
We showed that  the action of all the constraints restricted to 
 the connection part of the 
N patch data is some  combination of a 2d 
diffeomorphism and an SO(2,1) gauge rotation. Thus,
flat and null patches can only be 
moved around but never created or destroyed by the action of 
the constraints.
This meant that if we could actually 
build such solutions to the constraints in a way that 
holonomies of the connections around non-contractible loops 
in the flat sectors were freely specifiable, the dimensionality 
of the reduced phase space would be infinite. The rationale for 
building the particular example of
{\em initial data} for N patches in which the holonomies 
around loops in flat patches were freely specifiable, was to show 
that these types of solutions really do exist. It is very 
important to emphasize that we never had to fix any gauge,
(there is no BV gauge!) so it is impossible that we arrived at
any ``erroneous" conclusion as a consequence of the use of a 
partial gauge fixing procedure as Manojlovi\'c 
and Mikovi\'c claim. \\

For the sake of completeness we repeat now the arguments of
 \cite{VB}
explicitly in the context of \cite{MM1}. To this end we consider 
initial data of the type
\cite{VB},\cite{MM1}:
\begin{equation}
{\tilde E}^{\theta}_I=E_1 x_I,\hspace{5mm}A_{\theta}^I=A_1 x^I,
\hspace{5mm}{\tilde E}^{\phi}_I =E_2 y_I+E_3 t_I, \hspace{5mm}
A_{\phi}^I=A_2 y^I+A_3 t^I
\label{1}
\end{equation}
where $E_1$, $E_2$, $E_3$, $A_1$, $A_2$, $A_3$  are functions 
of $\theta$ (we coordinatize the torus with $\theta,\,\phi\in 
[0,2\pi)$ and identify 0 and $2\pi$) and $x_I$, $y_I$, $t_I$ 
are an orthonormal basis in the Lie algebra of $SO(2,1)$.
This ansatz {\it is not a gauge fixing condition}; its purpose
is only to select a subset of the possible initial data in order to 
study them in more detail. Inserting (\ref{1}) in the gauge
transformations generated by the 2+1 dimensional Ashtekar 
constraints 
with gauge parameters of 
the type \footnote{\noindent Notice that $\epsilon^1$, 
$\epsilon^2$, and, $\epsilon^3$ correspond just to a subset of all 
the possible gauge transformations generated by the Gauss law, vector 
and scalar constraint respectively. 
In general, $N^I, N^a, \NN$ are functions of $(\theta, \phi )$
with $N^I, N^a$ pointing in any direction.} 
\begin{equation}
N^I=\epsilon^1(\theta) x^I,\hspace{5mm}N^a \partial_a=
\epsilon^2(\theta)\partial_{\theta},\hspace{5mm}\NN=\epsilon^3
(\theta)
\label{2}
\end{equation}
we find equations 
\begin{eqnarray}
& & \delta A_1=-\frac{d\epsilon^1}{d\theta}+(E_2 f_3-E_3 f_2)
    \epsilon^3\label{3}\\
& & \delta A_2=-A_3\epsilon^1+f_2 \epsilon^2+E_1 f_3\epsilon^3
    \label{4}\\
& & \delta A_3=-A_2\epsilon^1+f_3\epsilon^2+E_1f_2\epsilon^3
    \label{5}
\end{eqnarray}
Note that due to the error described in {\bf (i)} above,
(\ref{5}) replaces the incorrect equation (19) of \cite{MM1}.

Let us consider the infinitesimal transformations on a null 
connection for which $f_2=f_3$ and $E_2=E_3$ (a similar analysis 
for 
$f_2= -f_3$ is possible). Consider $\epsilon^1
=\epsilon^2=0$ i.e. a gauge transformation generated by the
scalar constraint
\begin{eqnarray}
& & \delta A_1=0\nonumber\\
& & \delta A_2=E_1 f_3 \epsilon^3\label{9}\\
& & \delta A_3=E_1 f_2 \epsilon^3\nonumber
\end{eqnarray}
If, instead, we consider the transformations with 
$\epsilon^1=\epsilon^3=0$ (gauge transformations generated by the
vector constraint equivalent to diffeomorphisms modulo
$SO(2,1)$ transformations) we find
\begin{eqnarray}
& & \delta A_1=0\nonumber\\
& & \delta A_2=f_2\epsilon^2=f_3 \epsilon^2\label{10}\\
& & \delta A_3=f_3\epsilon^2=f_2\epsilon^2\nonumber
\end{eqnarray}
Thus, by choosing $\epsilon^2=E_1\epsilon^3$, the infinitesimal 
tranformations generated by the scalar constraint just reduce
to diffeomorphisms (modulo $SO(2,1)$ transformations). For a flat
connection this is true, as well, because we have now $\delta A_i=
0$, $i=1,2,3$. As a consequence we conclude that the action of the
scalar constraint smeared with a lapse $\epsilon^3$ is equivalent
to the action of the vector constraint with shift $\epsilon^2=
\epsilon^3 E_1$ as originally claimed in \cite{VB}. A direct
consequence of this is 
that no 
gauge transformation 
can create or destroy patches 
because it is impossible to do this by the action of any 
diffeomorphism and/or $SO(2,1)$ gauge rotation.

\noindent {\bf (iv)} Once the error in $\delta A_3$ is corrected, 
equation (46) 
in \cite{MM1} gives 
$$\delta W=-4\pi\frac{\sinh (\pi\sqrt{A_2^2-A_3^2})}{\sqrt{A_2^2-A_3^2}}
\left[(A_2f_2-A_3f_3)\epsilon^2+E_1(A_2f_3-A_3f_2)
\epsilon^3\right]$$
which is zero inside flat patches, in perfect agreement
 with the results in \cite{VB}.

In conclusion, we have shown that \cite{MM1} has errors and
does not point out any fallacy in our 
previous work \cite{VB}. There certainly exist interesting  subtle 
open issues which impinge on the validity/relevance of \cite{VB} 
(whether our data are in a singular part of phase space
(see \cite{VB2,VB3}), whether finite evolution could result in singular
$\tilde{E}^a_I$ etc.), but these are not the issues discussed in 
\cite{MM1}.

\end{document}